\documentstyle[preprint,aps,prb]{revtex}
\begin{document}
\preprint{ NSF-ITP-94-1}
\draft
\title{
Conserving approximations for the attractive Holstein and Hubbard models
}
\author{J. K. Freericks }
\address{
Department of Physics,
University of California,
Davis, CA 95616}
\date{\today}
\maketitle
\widetext
\begin{abstract}
Conserving approximations are applied to the attractive
Holstein and Hubbard models (on an infinite-dimensional hypercubic lattice).
All effects of nonconstant density of states and vertex corrections are
taken into account in the weak-coupling regime.  Infinite summation of
certain classes of diagrams turns out to be a quantitatively
less accurate approximation than truncation of the conserving
approximations to a finite order, but the infinite summation approximations do
show the correct qualitative behavior of generating a peak in the transition
temperature as the interaction strength increases.
\end{abstract}
\section{Introduction}
It is generally believed that the theoretical aspects of conventional
superconductors are well understood and that quantitative predictions
agree with experiment\cite{parks,reviews}.  The reason why low temperature
superconductors can be described accurately for all physical values of
the electron-phonon coupling is due to Migdal's theorem\cite{migdal}:
the ratio of the electron mass to the ion-core mass provides the small
parameter
that guarantees rapid convergence of the theory.  Eliashberg\cite{eliashberg}
generalized Migdal's theorem to the superconducting state and provided the
framework for quantitative calculations of the superconducting properties of
real materials\cite{parks,reviews}.

A more precise way of stating Migdal's theorem is to say that only the
electrons
that lie in an energy shell of width $\Omega_{\rm Debye}$ about the Fermi
surface are affected by phonon scattering,
and the only important scattering events involve the virtual
emission and reabsorption of phonons in an {\it ordered} fashion,
where the last emitted phonon is the first absorbed phonon, and so on.
Migdal-Eliashberg (ME) theory neglects
vertex corrections (which involve crossings of the phonon lines) and is an
accurate approximation for small phonon frequencies.  The remaining
unanswered question is how large does the phonon frequency have to be
before the effects of vertex corrections are observable?

There are many materials that are hypothesized to be electron-phonon
mediated superconductors, but have large phonon frequencies.
Ba$_{1-x}$K$_x$BiO$_3$
is a charge-density-wave (CDW) insulator at zero doping ($x=0$), but
becomes a superconductor (SC) away from half-filling\cite{BKBOexp} ($x\ge
0.37$).
The maximum phonon frequency is $\Omega_{\rm max.}=80~{\rm meV}$ for the
optical oxygen modes,\cite{BKBOphon} while the bandwidth\cite{BKBOband}
is $W=4~{\rm eV}$,
so the ratio of the phonon energy scale to the electronic energy scale is
$\Omega_{\rm max.}/W=0.02$.  Are vertex corrections important for this
material?

Alkali-metal-doped C$_{60}$ is another superconducting material that is
hypothesized to have an electron-phonon pairing mechanism.  There are
very high frequency phonons that correspond to distorting the C$_{60}$
cage\cite{C60phon} ($\Omega\approx 0.2~{\rm eV}$), while the electronic
bandwidth is quite narrow\cite{C60band} ($W=1.0~{\rm eV}$),
resulting in $\Omega_{\rm max.}/W\approx 0.2$.  Clearly, vertex corrections
must play an important role in phonon-mediated pairing mechanisms
in these materials.

The effect of vertex corrections on superconducting properties, in particular,
on the superconducting transition temperature, have been studied in the
past\cite{grabowski_sham,cai,freericks_scalapino}.  Grabowski and
Sham\cite{grabowski_sham} showed that vertex corrections lower $T_c$ for the
repulsive electron gas with $T_c\rightarrow 0$ for some critical value of
the plasma frequency.  The electron-phonon interaction has also been
examined,\cite{cai,freericks_scalapino} and, in general, vertex corrections
also
cause $T_c$ to drop.  Is this always the effect of vertex corrections, or can
vertex corrections sometimes cause an enhancement to $T_c$?

In this contribution the effects of vertex corrections are examined in a
systematic fashion via weak-coupling conserving approximations for the
attractive Holstein\cite{holstein} and Hubbard\cite{hubbard} models.
The effects of Coulomb repulsion are explicitly neglected here.
A detailed comparison of these perturbation schemes can be made to
exact results for these models in the infinite-dimensional limit to
determine which weak-coupling approximation is the most accurate.

The Holstein model consists of conduction electrons  that
interact with localized (Einstein) phonons:
\begin{equation}
  H = - {{t^*}\over{2\sqrt{d}}} \sum_{\langle j,k\rangle \sigma} (
   c_{j\sigma}^{\dag }
  c_{k\sigma} + c_{k\sigma}^{\dag  } c_{j\sigma} ) + \sum_j  (gx_j-\mu)
  (n_{j\uparrow}+n_{j\downarrow}-1) + {{1}\over{2}} M \Omega^2 \sum_j x_{j}^2 +
  {{1}\over{2}} \sum_j {p_{j}^2\over M}
\label{eq: holham}
\end{equation}
where $c_{j\sigma}^{\dag}$ ($c_{j\sigma}$) creates (destroys) an electron at
site $j$ with spin $\sigma$, $n_{j\sigma}=c_{j\sigma}^{\dag}c_{j\sigma}$ is
the electron number operator, and $x_j$ ($p_j$) is the phonon coordinate
(momentum) at site $j$.  The hopping matrix elements connect the nearest
neighbors of a hypercubic lattice in $d$-dimensions.  The
unit of energy is chosen to be the rescaled matrix
element $t^*$.  The phonon has a mass $M$ (chosen to be $M=1$), a
frequency $\Omega$, and a spring constant $\kappa\equiv M\Omega^2$ associated
with it.  The electron-phonon coupling constant (deformation potential)
is denoted by $g$; the effective electron-electron interaction strength is then
the bipolaron binding energy
\begin{equation}
  U\equiv - {{g^2}\over{M\Omega^2}}=-{g^2\over\kappa} \quad .
 \label{eq: udef}
\end{equation}
The chemical potential is denoted by $\mu$ and particle-hole symmetry occurs
for $\mu=0$.

In the instantaneous limit where $U$ remains finite and $g$ and $\Omega$ are
large compared to the bandwidth $(g,\Omega\rightarrow\infty ,U={\rm
finite})$, the Holstein model maps onto the attractive Hubbard
model\cite{hubbard}
\begin{equation}
  H = - {{t^*}\over{2\sqrt{d}}} \sum_{\langle j,k\rangle \sigma} (
   c_{j\sigma}^{\dag }
  c_{k\sigma} + c_{k\sigma}^{\dag  } c_{j\sigma} ) -\mu\sum_j (n_{j\uparrow}+
n_{j\downarrow}) + U \sum_j   (n_{j\uparrow}-{1\over 2})(n_{j\downarrow}-
{1\over 2})
\label{eq: hubbham}
\end{equation}
with $U$ defined by Eq.~(\ref{eq: udef}).

The weak-coupling theory is based upon the conserving approximations of
Baym and Kadanoff\cite{baym}: the free energy functional $\Phi$ is approximated
by a series expansion of skeleton diagrams of the dressed Green's function
$G$; the self energy $\Sigma(i\omega_n)$ is determined by functional
differentiation $\Sigma(i\omega_n)=\delta\Phi/\delta G(i\omega_n)$ at each
Matsubara frequency
$\omega_n\equiv (2n+1)\pi T$; and the irreducible vertex functions
$\Gamma(i\omega_m,i\omega_n)$ (in the relevant channels) are determined by
a second functional differentiation.

Independently, Van Dongen\cite{vandongen}  and Mart\'in-Rodero and
Flores\cite{martin_flores} showed that the free energy must be expanded to
order
$U^2$ to determine the correct transition temperature in the limit
$|U|\rightarrow 0$ for the Hubbard model at half-filling.  The vertex
corrections reduced the Hartree-Fock transition temperature by a
{\it factor of order
three}, but the gap ratio $2\Delta (0)/k_BT_c\approx 3.53$ was
unchanged (to lowest order).

In addition to reproducing the weak-coupling limit properly, one hopes that
the conserving
approximations will also be able reproduce the peak in the transition
temperature as a function of interaction strength that occurs as the
system crosses over from a weak-coupling regime (where pair formation
and condensation both occur at $T_c$) to a strong-coupling regime (where
preformed pairs order at a lower
temperature)\cite{nozieres_shmittrink,bipolarons}.  It will turn out that this
feature is not easily reproduced by a truncated conserving approximation.

The infinite-dimensional limit of Metzner and Vollhardt\cite{metzner_vollhardt}
is taken $(d\rightarrow\infty)$, in which the electronic many-body problem
becomes a local (impurity) problem that retains its complicated dynamics in
time.  The large-dimensional limit is quite useful because both the
Holstein\cite{freericks_jarrell_scalapino} and
Hubbard\cite{jarrell,kotliar_hub,georges_hub}
models can be solved exactly using the quantum Monte Carlo
(QMC) techniques of Hirsch and Fye\cite{hirsch_fye}.  These exact solutions
have
many of the qualitative features of the many-body problem in finite dimensions.
They also provide a
unique testing ground for various weak-coupling theories, since the approximate
theory can be compared directly to the exact solution in the thermodynamic
limit.

In the infinite dimensional limit,
the hopping integral is scaled to zero in such a fashion that the
free-electron kinetic energy remains finite while the self energy
for the single-particle Green's function and the irreducible vertex functions
have no momentum dependence and are functionals of the
local Green's function\cite{metzner_vollhardt,schweitzer_czycholl,metzner}.
This limit retains the strong-correlation effects that arise from
trying to simultaneously minimize both the kinetic energy
and the potential energy.

The many-body problem is solved by mapping it onto an auxiliary impurity
problem\cite{brandt_mielsch,okhawa} in a time-dependent field
that mimics the hopping of an electron onto
a site at time $\tau$ and off the site at a time $\tau '$.  The
action for the impurity problem is found by integrating out all of the
degrees of freedom of the other lattice sites in a path-integral
formalism.\cite{georges_kotliar}  The result is an effective action
\begin{eqnarray}
S_{eff.} &= \sum_{\sigma} \int_0^{\beta} d\tau \int_0^{\beta} d\tau '
c_{\sigma}^{\dag}(\tau)G_0^{-1}(\tau-\tau ')c_{\sigma}(\tau ')
+\sum_{\sigma}\int_0^{\beta}d\tau[gx(\tau)-\mu][c_{\sigma}^{\dag}(\tau)
c_{\sigma}(\tau)-1]\cr
&+ {1\over2}M\int_0^{\beta}d\tau [\Omega^2x^2(\tau)+\dot x^2(\tau)]
\label{eq: seff}
\end{eqnarray}
where $G_0^{-1}$ is the ``bare'' Green's function that contains
{\it all of the dynamical information of the other sites of the lattice}.
The interacting Green's function, defined to be
\begin{equation}
G(i\omega_n)\equiv -\int_0^{\beta} d\tau e^{i\omega_n\tau}
{{\rm Tr} \langle e^{-\beta H}T_{\tau} c(\tau)
c^{\dag}(0)\rangle\over {\rm Tr} \langle e^{-\beta H}\rangle } ~,
\label{eq: greendef}
\end{equation}
is determined by Dyson's equation
\begin{equation}
G_n^{-1}\equiv G^{-1}(i\omega_n) = G_0^{-1}(i\omega_n)-\Sigma (i\omega_n).
\label{eq: gdef}
\end{equation}

A self-consistency relation is required in order to determine the bare
Green's function $G_0$.  This is achieved by mapping the impurity problem
onto the infinite-dimensional lattice thereby equating the full Green's
function for the impurity problem with the local Green's function for
the lattice
\begin{equation}
G_{jj}(i\omega_n)=\sum_{\bf k} G({\bf k},i\omega_n) = \sum_{\bf k}
[i\omega_n+\mu-E({\bf k})-\Sigma(i\omega_n)]^{-1}
= F_{\infty}[i\omega_n+\mu -\Sigma(i\omega_n)].
\label{eq: gloc}
\end{equation}
Here $F_{\infty}(z)$ is the scaled complimentary error function of a complex
argument\cite{georges_kotliar}
\begin{equation}
F_{\infty}(z)\equiv {1\over{\sqrt{\pi}}}\int_{-\infty}^{\infty} dy
{\exp(-y^2)\over{z-y}} =-i{\rm sgn}[{\rm Im}(z)]\sqrt{\pi}e^{-z^2}{\rm erfc}
\{-i{\rm sgn}[{\rm Im}(z)]z\}.
\label{eq: fdef}
\end{equation}
The dynamics of the (local) impurity problem is identical to the dynamics
of the Anderson impurity
model\cite{schweitzer_czycholl,brandt_mielsch,okhawa,georges_kotliar,jarrell}
and is determined by employing a weak-coupling
conserving approximation for the local problem and satisfying the
self-consistency relation in Eq.~(\ref{eq: gloc}).

It is important to note that since one does not {\it a priori} know the
bare Green's function $G_0^{-1}$ in Eq.~(\ref{eq: seff}), one must iterate
to determine a self-consistent solution for the Green's
function of the infinite-dimensional lattice.  This is done by performing
self-consistent perturbation theory for the self energy $\Sigma[G]$ within
a conserving approximation, and then determining the new local Green's
function from the approximate self energy and
Eq.~(\ref{eq: gloc}).  This process is iterated until
convergence is achieved
[the maximum variation of each $G(i\omega_n)$ is less than one part in
$10^8$ which typically takes between 5 and 30 iterations].

Static two particle properties are also easily calculated since the
irreducible vertex function is local\cite{zlatic_horvatic}.  The
static susceptibility for CDW order is given by
\begin{eqnarray}
\chi^{CDW}({\bf q})&\equiv& {1\over2N}\sum_{{\bf R}_j-{\bf R}_k\sigma\sigma'}
e^{i{\bf q}\cdot
({\bf R}_j-{\bf R}_k)} T\int_{0}^{\beta} d\tau \int_{0}^{\beta} d\tau '
[\langle n_{j\sigma}(\tau) n_{k\sigma'}(\tau ')\rangle -
\langle n_{j\sigma}(\tau)\rangle\langle n_{k\sigma'}
(\tau ')\rangle ] \cr
&\equiv& T\sum_{mn} \tilde\chi^{CDW} ({\bf q},i\omega_m,i\omega_n)=
T\sum_{mn} \tilde\chi_{mn}^{CDW} ({\bf q})\quad,
\label{eq: chicdw}
\end{eqnarray}
at each ordering wavevector ${\bf q}$.  Dyson's equation for the two-particle
Green's function becomes\cite{jarrell,zlatic_horvatic}
\begin{equation}
\tilde\chi_{mn}^{CDW}({\bf q})=\tilde\chi_{m}^0({\bf q})\delta_{mn}
-T\sum_p \tilde\chi_m^0({\bf q})\Gamma_{mp}^{CDW}\tilde
\chi_{pn}^{CDW}({\bf q})\quad ,
\label{eq: cdwdys}
\end{equation}
with $\Gamma_{mn}^{CDW}$ the (local) irreducible vertex function in the CDW
channel.

The
bare CDW susceptibility $\tilde\chi_n^0({\bf q})$ in
Eq.~(\ref{eq: cdwdys}) is defined in terms of the single-particle Green's
function
\begin{eqnarray}
\tilde\chi_n^0({\bf q})&\equiv&-{1\over N} \sum_{\bf k} G_n({\bf
k})G_n({\bf k+q})\cr
&=&-{1\over{\sqrt{\pi}}}{1\over{\sqrt{1-X^2({\bf q})}}}\int_{-\infty}^{\infty}
dy {{e^{-y^2}}\over{i\omega_n+\mu-\Sigma_n-y}}F_{\infty}\left [ {{i\omega_n+
\mu-\Sigma_n-X({\bf q})y}\over{\sqrt{1-X^2({\bf q})}}}\right ]
\label{eq: chi0cdw}
\end{eqnarray}
and all of the wavevector dependence is included in the
scalar\cite{brandt_mielsch,muellerhartmann} $X({\bf q})
\equiv \sum\nolimits_{j=1}^d \cos {\bf q}_j/d$.  The mapping ${\bf q}
\mapsto X({\bf q})$ is a many-to-one mapping that determines an equivalence
class of wavevectors in the Brillouin zone.  ``General'' wavevectors are
all mapped to $X=0$ since $\cos {\bf q}_j$ can be thought of as a random
number between $-1$ and 1 for ``general'' points in the Brillouin zone.
Furthermore, all possible values of $X$ $(-1\le X\le 1)$ can be labeled
by a wavevector that lies on the diagonal of the first Brillouin zone extending
from the zone center $(X=1)$ to the zone corner $(X=-1)$.  The presence
of incommensurate order in the attractive Holstein model is restricted to a
very
narrow region of parameter space\cite{freericks_jarrell_scalapino,feinberg}
so only the ``antiferromagnetic'' point $X=-1$ is considered for CDW order.
The integral for $\tilde\chi_m^0(X)$
in Eq.~(\ref{eq: chi0cdw}) can then be
performed analytically\cite{brandt_mielsch}
$\tilde\chi_n^0(X=-1)=-{{G_n}/({i\omega_n+\mu-\Sigma_n}})$.
The irreducible vertex function $\Gamma_{mn}^{CDW}$ is calculated
perturbatively
from the dressed Green's functions in a conserving approximation (see below).

A similar procedure is used for the singlet $s$-wave SC channel.  The
corresponding definitions are as follows:  The static susceptibility
in the superconducting channel is defined to be
\begin{eqnarray}
\chi^{SC}({\bf q})&\equiv& {1\over N}\sum_{{\bf R}_j-{\bf R}_k}
e^{i{\bf q}\cdot
({\bf R}_j-{\bf R}_k)} T\int_{0}^{\beta} d\tau \int_{0}^{\beta} d\tau '
\langle c_{j\uparrow}(\tau)
c_{j\downarrow}(\tau)c_{k\downarrow}^{\dag}(\tau ')c_{k\uparrow}^{\dag}(\tau ')
\rangle  \cr
&\equiv& T\sum_{mn} \tilde\chi^{SC} ({\bf q},i\omega_m,i\omega_n)=
T\sum_{mn} \tilde\chi_{mn}^{SC} ({\bf q})\quad,
\label{eq: chisc}
\end{eqnarray}
for superconducting pairs that carry momentum ${\bf q}$;  Dyson's equation
becomes
\begin{equation}
\tilde\chi_{mn}^{SC}({\bf q})=\tilde\chi_{m}^0{'}({\bf q})\delta_{mn}
-T\sum_p \tilde\chi_m^0{'}({\bf q})\Gamma_{mp}^{SC}\tilde
\chi_{pn}^{SC}({\bf q})\quad ,
\label{eq: scdys}
\end{equation}
with $\Gamma_{mn}^{SC}$ the corresponding irreducible vertex function for the
SC channel; the bare pair-field susceptibility becomes
\begin{eqnarray}
\tilde\chi_n^0{'}({\bf q})&\equiv&{1\over N} \sum_{\bf k} G_n({\bf
k})G_{-n-1}({\bf -k+q})\cr
&=&{1\over{\sqrt{\pi}}}{1\over{\sqrt{1-X^2({\bf q})}}}\int_{-\infty}^{\infty}
dy {{e^{-y^2}}\over{i\omega_n+\mu-\Sigma_n-y}}F_{\infty}\left [
{{i\omega_{-n-1}+
\mu-\Sigma_{-n-1}-X({\bf q})y}\over{\sqrt{1-X^2({\bf q})}}}\right ]
\label{eq: chi0sc}
\end{eqnarray}
with the special value
$\tilde\chi_n^0{'}(X=1)=G_n/(i\omega_{-n-1}+\mu-\Sigma_{-n-1})$ for the
SC pair that carries no net momentum; and finally the irreducible vertex
function is also determined in the conserving formalism (see below).

At this point the transition temperature of the infinite-dimensional Holstein
model is found by calculating the temperature at which the relevant
susceptibility diverges (CDW or SC).

Section II contains the comparison of QMC exact solutions to ME theory and
the second-order conserving approximation for the Holstein model.
Analytical expressions
for the change in $T_c$ due to vertex corrections are given for
the SC channel.  Section III includes the application of conserving
approximations to the attractive Hubbard model at half-filling.  Truncated
conserving approximations through fourth order are compared to the different
fluctuation-exchange approximations and the exact QMC solutions.
Conclusions are presented in Section IV.

\section{Holstein Model}

There are two different types of approximations that are generally made for
the electron-phonon interaction:  the first method is a truncated conserving
approximation that includes all vertex corrections to a finite order and
is valid for all values of the phonon
frequency\cite{baym,freericks_jarrell_scalapino,feinberg,freericks_scalapino};
the second method is ME theory in which vertex corrections are neglected,
but the phonon propagator is dressed to all
orders\cite{migdal,eliashberg,marsiglio}.  These two methods are compared and
contrasted in Figures 1 and 2.  Figure 1 (a) shows the self energy for a
conserving approximation through second order.  The self energy includes,
respectively, the Hartree term (which is a constant and can be reabsorbed into
the chemical potential), the Fock term, the second-order term that dresses
the phonon propagator, and the lowest-order vertex correction.  Figure 1 (b)
displays the corresponding self-consistent equations for ME theory:
the self energy includes the Hartree term (which can once again be reabsorbed
into the chemical potential) and the Fock term (which is calculated with
the {\it dressed} phonon propagator\cite{marsiglio}).  The dressed phonon
propagator satisfies Dyson's equation [Figure 1 (c)].

To be more explicit, the self energy for the second-order conserving
approximation is
\begin{equation}
\Sigma_n(cons.)=-g^2T\sum_rD_{n-r}G_r+g^4T^2\sum_{rs}[-2D_{n-r}+D_{r-s}]D_{n-r}
G_rG_sG_{n-r+s}\quad ,
\label{eq: sigma_cons}
\end{equation}
which includes the Fock diagram contribution and the two second-order
contributions in Figure 1 (a).  The bare phonon propagator
$D_l\equiv D(i\omega_l)$ in Eq. (\ref{eq: sigma_cons}) is given by
\begin{equation}
D_l=-{1\over M(\Omega^2+\omega_l^2)}\quad ,
\label{eq: ddef}
\end{equation}
for each (bosonic) Matsubara frequency $\omega_l\equiv 2l\pi T$.
On the other hand, in ME theory, the self energy satisfies
\begin{equation}
\Sigma_n(ME) =g^2T\sum_r{G_r\over M(\Omega^2+\omega_{n-r}^2)+\Pi_{n-r}}
\quad;\quad
\Pi_l\equiv 2g^2T\sum_r G_{l+r}G_r
\label{eq: sigma_me}
\end{equation}
with $\Pi_l$ the phonon self energy (evaluated in the limit where vertex
corrections are neglected).

The self-consistency step involves determining a new local Green's function
$G_n$ from the integral relation in Eq.~(\ref{eq: gloc}) with the
approximate self-energy of Eq.~(\ref{eq: sigma_cons}) or
Eq.~(\ref{eq: sigma_me}).
This process is repeated until the maximum deviation in the local Green's
function is less than one part in $10^8$.

Once the Green's functions and self energies have been determined, the
irreducible vertex functions can be calculated for the CDW or SC
channels.  The vertices are simple in the ME theory: the irreducible
vertex in the CDW channel satisfies [see Figure 2 (a)]
\begin{equation}
\Gamma_{mn}^{CDW}(ME)=2U\quad;
\label{eq: cdw_me}
\end{equation}
and the irreducible vertex function in the SC channel satisfies [see Figure 2
(b)]
\begin{equation}
\Gamma_{mn}^{SC}(ME)=-g^2{1\over M(\Omega^2+\omega_{m-n}^2)+\Pi_{m-n}}\quad .
\label{eq: sc_me}
\end{equation}
The phonon propagator in the CDW vertex is the {\it bare} propagator to
avoid a double counting of diagrams.

The irreducible vertex functions acquire more structure in the second-order
conserving approximation.  In the CDW channel [see Figure 2 (c)] one must
include both direct and exchange diagrams as well as the vertex corrections.
The result is
\begin{eqnarray}
\Gamma_{mn}^{CDW}(cons.)&=&2U-2U^2T\sum_r[G_rG_{m-n+r}+G_rG_{m+n-r}]
[{\Omega^2\over \Omega^2+\omega_{n-r}^2}]^2\cr
&-&U{\Omega^2\over \Omega^2+\omega_{m-n}^2}-2U^2T\sum_rG_rG_{m-n+r}
{\Omega^2\over \Omega^2+\omega_{m-n}^2}[{\Omega^2\over \Omega^2+\omega_{m-n}^2}
-{\Omega^2\over \Omega^2+\omega_{n-r}^2}]\cr
&+&U^2T\sum_rG_rG_{m+n-r}{\Omega^2\over \Omega^2+\omega_{m-r}^2}{\Omega^2\over
\Omega^2+\omega_{n-r}^2}\quad .
\label{eq: cdw_cons}
\end{eqnarray}
Note that the vertex corrections (arising from the first-order exchange
diagrams) modify the {\it interaction} in the CDW channel so that it properly
interpolates between the zero frequency limit $\Gamma^{CDW}\rightarrow 2U$
and the infinite frequency limit $\Gamma^{CDW}\rightarrow U$.  At an
intermediate
frequency, the CDW interaction strength has a complicated temperature
dependence.  In the SC channel [see Figure 2 (d)] one finds
\begin{eqnarray}
\Gamma_{mn}^{SC}(cons.)&=&U {\Omega^2\over \Omega^2+\omega_{m-n}^2}\cr
&+&U^2T\sum_r
G_rG_{m-n+r}{\Omega^2\over \Omega^2+\omega_{m-n}^2}[2{\Omega^2\over\Omega^2
+\omega_{m-n}^2}-{\Omega^2\over\Omega^2+\omega_{n-r}^2}-
{\Omega^2\over\Omega^2+\omega_{m+r+1}^2}]\cr
&-&U^2T\sum_rG_rG_{m+n+r+1}{\Omega^2\over\Omega^2+\omega_{m+r+1}^2}
{\Omega^2\over\Omega^2+\omega_{n+r+1}^2}\quad .
\label{eq: sc_cons}
\end{eqnarray}

As the transition temperature (to a CDW-ordered state or a SC-ordered state)
is approached from above, the susceptibility (in the relevant channel)
diverges.  Therefore, one can determine the transition temperature by
finding the temperature where the scattering matrix (in the relevant
channel)
\begin{equation}
T_{mn}=-T\Gamma_{mn}\chi_n^0\quad,
\label{eq: Tmat}
\end{equation}
has unit eigenvalue\cite{owen_scalapino}.  In general, the eigenvector
corresponding to the maximum eigenvalue of the scattering matrix is
symmetric with respect to a change in sign of the Matsubara frequency.

At half filling the Holstein model interaction is particle-hole symmetric,
so the Green's functions and self energies are purely imaginary and the
vertices
are real.  The self energy can be expressed by $\Sigma(i\omega_n)\equiv
i\omega_nZ(i\omega_n)$, with $Z(i\omega_n)$ the renormalization function
for the self energy.  At half filling, both $\chi_m^0(X=-1)$ and
$\chi_m^{0 \prime} (X=1)$ are also even
functions of the Matsubara frequency, so the only contribution of the
irreducible vertex function to the eigenvalue of the scattering matrix
comes from the even Matsubara frequency
component $[\Gamma_{m,n}+\Gamma_{-m-1,n}]/2$.

In order to judge the accuracy of these approximate methods for the electronic
self energy and the irreducible vertex functions, it is necessary to compare
them to the exact results.  The best way to do this would be to directly
compare
the perturbative results to the exact QMC results.  Unfortunately, there is
no available QMC data to do this.  However, it has been well established,
that the iterated perturbation theory (IPT) of Georges and
Kotliar\cite{georges_kotliar} yields essentially exact results for the
electronic self energy of the Hubbard model (by direct comparison with the QMC
results\cite{georges_hub}) as long as the system is at half filling.  The IPT
is identical to the second-order conserving approximation, except that the
perturbation theory is strictly truncated to second order in $U$.

A comparison of the approximations to the IPT results for the electronic
self energy and a comparison of the approximations for
one column of the irreducible vertex function in the CDW channel
is made in Figures 3 and 4 for two different interaction
strengths at half filling.  The phonon frequency is set to be approximately
one-eighth of the effective bandwidth ($\Omega/t^*=0.5$) as was done in the
QMC solutions\cite{freericks_jarrell_scalapino}.  The energy cutoff is set to
include 256 positive Matsubara frequencies for the perturbative approximations.

At weak coupling $(g=0.4$, Figure 3), the second-order conserving approximation
clearly provides a more accurate approximation to the electron self energy
(under the assumption that the IPT is essentially exact).  One expects the
exact irreducible vertex function to be a frequency dependent interaction,
so the second-order conserving approximation is probably more accurate here
too [the CDW vertex for the ME theory has
no frequency dependence as shown in Eq.~(\ref{eq: cdw_me})].

An underestimation of the self energy causes an overestimation of the
transition temperature and vice versa.  Similarly, an underestimation of the
magnitude of the irreducible vertex will cause an underestimation of $T_c$
(and vice versa).  Since the ME theory overestimates the self energy
and overestimates the CDW vertex (because there is no weakening of the
CDW vertex at small frequency transfers), it is difficult to predict whether
or not ME theory will overestimate $T_c$.  In the same fashion we do not know
whether or not the second-order conserving approximation will overestimate
$T_c$ since the magnitude of the exact vertex is not known.

As the coupling strength is increased to the point where a double-well
structure
began to develop in the effective phonon potential of the QMC
simulations\cite{freericks_jarrell_scalapino} ($g=0.5$, Figure 4) one can see
strong-coupling effects begin to enter.  Surprisingly, ME theory is becoming a
more accurate approximation to the self energy here.
Stated in other words, as the coupling strength
increases, the effect of vertex corrections is reduced\cite{cai}.
Unfortunately, the self-consistent equations for the ME theory become
unstable to an iterative solution as the coupling strength is increased
further.

At half filling, the Holstein model always has a transition to a CDW-ordered
phase at ${\bf q}=(\pi ,\pi ,\pi ...)$ $(X=-1)$.  The transition temperature
to this commensurate CDW is plotted in Figure 5 as a function of the
interaction strength.  The second-order conserving approximation is compared
to ME theory and the QMC simulations\cite{freericks_jarrell_scalapino}.
The conserving approximation is much
more accurate at weak
coupling\cite{vandongen,martin_flores,freericks_scalapino} (ME theory
predicts a transition temperature that is an order of magnitude higher
than the QMC and conserving approximation results at the lowest value of the
coupling considered) because the inclusion of the first-order exchange
diagrams produces the correct interaction and the inclusion of the second-order
terms produces the correct prefactor.  However, ME theory does display the
proper
qualitative behavior of developing a peak in $T_c$ as the interaction
strength increases.  This feature is not reproduced by the truncated conserving
approximation.

As the system is doped away from half-filling, the CDW instability remains
locked at the commensurate point $(X=-1)$ until it gives way
to a SC instability (incommensurate order may appear in a very narrow region
of phase space near the CDW-SC phase
boundary\cite{freericks_jarrell_scalapino,feinberg}
but is neglected here).  In Figure 6, the phase diagram of the
Holstein model is plotted for two values of the interaction strength
($g=0.4$, $g=0.5$).  The weak-coupling QMC data ($g=0.4$) are reproduced most
accurately by the second-order conserving approximation, as expected from the
comparison of the self energy and the vertices in Figure 3.
The SC transition is reproduced
remarkably well, because the underestimation of the self energy [Figure 3 (a)]
must be compensated by an underestimation of the SC vertex. The
critical concentration for the CDW-SC phase boundary is also more accurately
determined by the conserving approximation.  Note that the difference
between the SC transition temperature calculated with ME theory and the
second-order conserving approximation explicitly shows the lowest-order
effect of vertex corrections.  The vertex corrections lower $T_c$ by about
a factor of two at the phase boundary, but are reduced as the doping increases.

At the stronger coupling strength $(g=0.5)$ ME theory reproduces the CDW
transition temperature very accurately at half filling, but not the doping
dependence of $T_c$.  It does manage to reproduce the transition temperature
in the SC sector quite well, but the second-order conserving approximation
is superior at determining the CDW-SC phase boundary.  Clearly both
approximation methods are inadequate at this large a value of the coupling
strength.

Up to this point we have concentrated on one value of the phonon frequency
and have compared the numerical solution of the self-consistent perturbation
theory with the numerically exact QMC solutions.  In the limit of weak coupling
$(|U|\rightarrow 0)$, the transition temperature approaches zero
$(T_c\rightarrow 0)$ and the leading behavior of $T_c$ can be determined
analytically\cite{vandongen,martin_flores,freericks_scalapino,feinberg}.
We will
concentrate on the SC channel only, because the analytical techniques are
not as accurate for the CDW channel.

In the limit $T\rightarrow 0$, the self energy satisfies
\begin{equation}
\lim_{T\rightarrow 0} [i\omega_n+\mu-\Sigma(i\omega_n)] = \bar\mu+i\omega_nZ
\quad ,
\label{eq: sigmalimit}
\end{equation}
where the renormalized chemical potential is
\begin{equation}
\bar\mu=\mu-\lim_{T\rightarrow 0} {\rm Re}\Sigma(i\omega_n)=\mu-U\rho_e+O(U^2)
\quad ,
\label{eq: mubardef}
\end{equation}
and the renormalization function is
\begin{equation}
Z=Z(0)=1+|U|\int_0^{\infty} dy {\rho(y+\mu)+\rho(-y+\mu)\over 2} {\Omega\over
(\Omega+y)^2} + O(U^2)\quad .
\label{eq: zdef}
\end{equation}
Here $\rho(y)\equiv\exp(-y^2)/\sqrt{\pi}$ is the noninteracting density of
states (DOS) in infinite dimensions.  The irreducible vertex in the SC
channel [Eq.~(\ref{eq: sc_cons})] becomes
\begin{equation}
\Gamma_{mn}^{SC}=-{\Omega^2\over\Omega^2+\omega_{m-n}^2}|U|[1+2|U|I_1]
+U^2I_2+O(U^3)\quad ,
\label{eq: sc_cons_0}
\end{equation}
with $I_1$ and $I_2$ two smooth temperature-dependent integrals that can be
approximated by their zero-temperature limit:
\begin{equation}
I_1\equiv-{1\over 2\pi}\int_{-\infty}^{\infty}dyF_{\infty}^2(iy+\mu)
{y^2\over\Omega^2+y^2}+O(U)\quad ;
\label{eq: I_1def}
\end{equation}
\begin{equation}
I_2\equiv-{1\over 2\pi}\int_{-\infty}^{\infty}dyF_{\infty}^2(iy+\mu)
{\Omega^4\over(\Omega^2+y^2)^2}+O(U)\quad .
\label{eq: I_2def}
\end{equation}
The bare susceptibility becomes
\begin{equation}
\chi_n^{0\prime}(X=1)=-{{\rm Im}
F_{\infty}(i\omega_nZ+\bar\mu)\over\omega_nZ}+O(U^2)
\quad ,
\label{eq: chi_0limit}
\end{equation}
for the electron pairs that carry no net momentum.

In the square-well approximation\cite{bcs,reviews}, the smooth temperature
dependence of the SC vertex is replaced by a sharp cutoff at a characteristic
frequency $\omega_c$:
\begin{equation}
{\Omega^2\over\Omega^2+\omega_{m-n}^2}\rightarrow \theta(\omega_c-|\omega_m|)
\theta(\omega_c-|\omega_n|)\quad ,
\label{eq: wcdef}
\end{equation}
with $\theta(x)$ the unit step function.  The scattering matrix
[in Eq.~(\ref{eq: Tmat})] then becomes
\begin{equation}
T_{mn}=-T[\theta(\omega_c-|\omega_m|)\theta(\omega_c-|\omega_n|)|U|(1+2|U|I_1)-
U^2I_2]{{\rm Im} F_{\infty}(i\omega_nZ+\bar\mu)\over\omega_nZ}\quad .
\label{eq: Tmatlimit}
\end{equation}
The SC transition temperature is now determined by solving the matrix
eigenvalue
equation $\sum_nT_{mn}\phi_n=\phi_m$.

The eigenvector $\phi_m$, can be chosen to be of the form
$\phi_m=1+a\theta(\omega_c-|\omega_m|)$ in the square-well approximation, so
that the matrix eigenvalue equation is reduced to two coupled algebraic
equations
\begin{eqnarray}
a&=&{\rho(\mu)|U|\over Z}(1+2|U|I_1)(1+a)R\quad ,\cr
1&=&-{\rho(\mu)U^2\over Z}I_2(aR+S)\quad ,
\label{eq: coupledeqns}
\end{eqnarray}
with $R$ and $S$ defined by
\begin{equation}
R\equiv-{T\over\rho(\mu)}\sum_{|\omega_n|<\omega_c}
{{\rm Im}F_{\infty}(i\omega_nZ+\bar\mu)\over\omega_n}\quad ,
\label{eq: rdef}
\end{equation}
\begin{equation}
S\equiv-{T\over\rho(\mu)}\sum_{n=-\infty}^{\infty}
{{\rm Im}F_{\infty}(i\omega_nZ+\bar\mu)\over\omega_n}\quad .
\label{eq: sdef}
\end{equation}

The infinite summation over Matsubara frequencies can be performed in the
standard fashion\cite{reviews} to yield
\begin{eqnarray}
S&=&{1\over 2}\int_{-\infty}^{\infty} {dy\over y} {\rho(y+\bar\mu)\over
\rho(\mu)}\tanh {y\over 2ZT_c}\quad ,\cr
&=&\ln{1\over 2ZT_c}+\int_0^{\infty}{dy\over y}(\tanh y-1+
{\rho(y+\bar\mu)+\rho(-y+\bar\mu)\over 2\rho(\mu)})\quad ,
\label{eq: Sintegral}
\end{eqnarray}
while the truncated summation can be expressed in an integral
form\cite{grabowski_sham}
\begin{eqnarray}
R&=&S-{1\over 2\pi}\int_{\omega_c}^{\infty}{dy\over y}{\rm Im}
F_{\infty}(iyZ+\bar\mu)\quad ,\cr
&=&S-{1\over \pi}\int_0^{\infty}{dy\over y}
{\rho(y+\bar\mu)+\rho(-y+\bar\mu)\over 2\rho(\mu)}\tan^{-1}{y\over Z\omega_c}
\label{eq: Rintegral}
\end{eqnarray}
if the transition temperature is much less than the cutoff frequency
$(T_c << \omega_c)$.

The coupled algebraic equations [in Eq.~(\ref{eq: coupledeqns})] are then
solved
by
\begin{equation}
{Z\over\rho(\mu)|U|}=[1+|U|(2I_1-I_2)]S+(1+2|U|I_1)(R-S)\quad ,
\label{eq: Tceqn}
\end{equation}
to order $|U|$.  The limiting form of the transition temperature is now
found by substituting Eqs.~(\ref{eq: Sintegral}) and (\ref{eq: Rintegral})
for $R$ and $S$ into Eq.~(\ref{eq: Tceqn}) and solving for $T_c$.  The
result is
\begin{equation}
T_c=\exp (-{1\over\rho(\mu)|U|}) f_{phonon}f_{electron}f_{DOS}f_{vertex}\quad ,
\label{eq: Tcfinal}
\end{equation}
which includes the interaction term and the constant prefactors.  The constant
terms arise from the phonon self energy, the electron self energy, the
nonconstant DOS, and the vertex corrections.

The phonon self energy correction is
\begin{equation}
f_{phonon}=\exp[-{1\over \pi\rho(\mu)}{\rm Re}
\int_{-\infty}^{\infty}dyF_{\infty}^2(iy+\mu)]=\exp
[\sqrt{2}\int_{-\infty}^{\infty}
{dw\over w}{\rho(w)\over\rho(\mu)}{\rm erf}(w+\sqrt{2}\mu)]\quad,
\label{eq: fphonon}
\end{equation}
and is independent of the phonon frequency.  This correction factor is normally
included in the definition of the electron-phonon interaction strength
$\lambda$
\begin{equation}
\lambda\equiv\rho(\mu)|U|[1-{|U|\over \pi}{\rm Re}\int_{-\infty}^{\infty}dy
F_{\infty}^2(iy+\mu)]\quad ,
\label{eq: lambda}
\end{equation}
of the ME theory formalism.

The electron self energy factor satisfies
\begin{equation}
f_{electron}=\exp [-\int_0^{\infty}dy e^{-y^2}\cosh
(2\mu y){\Omega\over(\Omega+y)^2}]\quad ,
\label{eq: felectron}
\end{equation}
which approaches the standard ME theory result of
$f_{electron}\rightarrow e^{-1}$ as $\Omega\rightarrow 0$ and approaches the
standard Hubbard model result of $f_{electron}\rightarrow 1$ as
$\Omega\rightarrow \infty$.

The nonconstant DOS factor is
\begin{equation}
f_{DOS}={1\over 2}\exp \int_0^{\infty} {dy\over y} [\tanh y - 1 +e^{-y^2}\cosh
(2\mu y)(1-{2\over \pi}\tan^{-1}{y\over\omega_c})]\quad ,
\label{eq: fdos}
\end{equation}
which depends on the square-well cutoff frequency $\omega_c$.  In the limit
$\omega_c\rightarrow 0$, $f_{DOS}$ approaches the ME theory result of $1.14
\omega_c$, whereas in the limit $\omega_c\rightarrow\infty$, $f_{DOS}$
approaches the Hubbard model result\cite{feinberg}
\begin{equation}
\lim_{\omega_c\rightarrow\infty}
f_{DOS}=0.85\exp[2\int_0^{\infty}{dy\over y} e^{-y^2}\sinh^2\mu y]
=0.85\exp[\sqrt{\pi}\int_0^{\mu}dye^{y^2}{\rm erf}(y)]\quad .
\label{eq: fdoslimit}
\end{equation}
Note that once the phonon energy scale is larger than the electronic energy
scale, it is the bandstructure (not $\omega_c$) that determines the DOS
prefactor.

Finally, the vertex correction factor becomes
\begin{equation}
f_{vertex}=\exp [{1\over\pi\rho(\mu)}{\rm Re}\int_{-\infty}^{\infty}dy
F_{\infty}^2
(iy+\mu){\Omega^2\over\Omega^2+y^2}(1+{1\over 2}{\Omega^2\over \Omega^2+y^2})]
\quad ,
\label{eq: fvertex}
\end{equation}
which approaches 1 as $\Omega\rightarrow 0$.  In the high-frequency limit, the
vertex corrections cancel the phonon self energy corrections and yield
\begin{equation}
\lim_{\Omega\rightarrow\infty}f_{phonon}f_{vertex}=\exp[-{1\over\sqrt{2}}
\int_{-\infty}^{\infty}
{dw\over w} {\rho(w)\over\rho(\mu)}{\rm erf}(w+\sqrt{2}\mu)]\quad ,
\label{eq: fphononvertex}
\end{equation}
which reproduces van Dongen's result\cite{vandongen} $f_{phonon}f_{vertex}=
\exp [-\sqrt{2}\ln (1+\sqrt{2})]$ at half filling $(\mu=0)$.  In the infinite
frequency limit (attractive Hubbard model), the vertex corrections always
reduce $T_c$.

The effect of vertex corrections upon the superconducting transition
temperature in the weak-coupling limit are displayed in Figure 7~(a).
The vertex correction factor, $f_{vertex}=T_c({\rm vertex})/T_c({\rm
no~vertex})$, is
plotted against the phonon frequency for eight different electron
concentrations.  At half filling, the vertex corrections sharply reduce
$T_c$, so that $T_c$ calculated with vertex corrections is a factor
of two lower than $T_c$
calculated without vertex corrections at $\Omega=0.13t^*$ (or, since the
effective electronic bandwidth is approximately $W=4t^*$, when
$\Omega/W=0.03$).
Therefore, vertex corrections should play an important role in
Ba$_{1-x}$K$_x$BiO$_3$ where $\Omega/W=0.02$.

As the system is doped away from
half filling the effect of the vertex corrections is reduced (as was already
seen in Figure 6), until a critical electron concentration
$(\rho_c\approx 0.2$) is reached where the vertex corrections initially
cause an {\it enhancement} to $T_c$.
This enhancement occurs because the electronic Green's functions have a
larger real part than imaginary
part which causes the integrand in Eq.~(\ref{eq: fvertex}) to change sign for
small $y$.  This enhancement will not be seen in standard ME theory with
a constant DOS, because the Green's functions are chosen to be purely imaginary
in that case.  As the phonon frequency is increased to a large enough value,
the vertex corrections will once again reduce $T_c$, because they always cause
a reduction in the limit $\Omega\rightarrow\infty$ [see
Eq.~(\ref{eq: fphononvertex})].

The square-well cutoff frequency, $\omega_c$, should vanish as the phonon
frequency vanishes, and should become infinite as the phonon frequency
becomes infinite. The cutoff frequency is chosen to be
three-fifths of the phonon frequency $(\omega_c=0.6\Omega)$, so that
the proper limiting behavior is
attained\cite{reviews,freericks_scalapino,vandongen} as $\Omega\rightarrow 0$
and $\Omega\rightarrow \infty$.  The prefactor to the SC transition
temperature in Eq.~(\ref{eq: Tcfinal}) is plotted in Figure 7~(b) for eight
different electron concentrations.  Note that there is an optimal phonon
frequency where the SC response is maximal, that shifts from a low frequency
at half filling to higher frequencies as the electron concentration is reduced.
The optimal phonon frequency is usually smaller than the effective bandwidth
($W=4t^*$).

In general, one must expand the free energy thru second
order\cite{vandongen,martin_flores}
to properly determine $T_c$ in the limit $|U|\rightarrow 0$.  The miracle of
ME theory is that a first-order calculation with dressed phonons properly
determines $T_c$ in the limit $\Omega\rightarrow 0$.  In the CDW channel, the
vertex corrections modify both the interaction strength and the prefactor,
while
in the SC channel, only the prefactor is modified.  It is this robustness
of the SC channel to the effects of vertex corrections that explains the
success of ME theory for low temperature superconductors.

\section{Hubbard Model}

The Hubbard model in Eq. (\ref{eq: hubbham}) is the infinite-frequency
limit $(\Omega\rightarrow\infty)$ of the Holstein model.  The Hubbard model
has an electron-electron interaction that only occurs between electrons
with opposite spins.  This happens because of the cancelation of the
direct and exchange diagrams which causes all electron-electron interactions
between like-spin particles to vanish.  The perturbation theory becomes
much simpler in the Hubbard model case, because of this reduction of
diagrams, and can be performed to higher order.  Here the truncated
conserving approximation will be carried out to fourth order, and will be
compared to the fluctuation-exchange (FLEX) approximation\cite{baym} to
determine the best way to approximate the Hubbard model in the
infinite-dimensional limit.  Previous work has concentrated on second-order
conserving approximations\cite{muellerhartmann,hirashima},
third-order conserving approximations\cite{freericks_scalapino}, or the FLEX
approximation\cite{menge_muellerhartmann}.

One expects that a truncated approximation will be superior to an infinite
summation of random-phase approximation (RPA) bubbles and particle-hole
and particle-particle ladders because the many-body problem reduces to
a self-consistently embedded Anderson impurity model, and the analysis
of Yamada\cite{yamada} has shown that the total fourth-order corrections to the
self energy are an order of magnitude smaller and opposite in sign to the
fourth-order contribution of the FLEX approximation.  The irreducible
vertex functions should have similar effects, but have not yet been analyzed
in detail.

The diagrammatic expansion for self energy (in a conserving approximation)
of the Hubbard model is given in Figure 8.  The first line includes the
first-order Hartree contribution (which is a constant that can be absorbed
into a renormalized chemical potential), the second-order contribution
and the third-order particle-hole and particle-particle ladders.  The
second line contains the fourth-order contributions from the RPA bubbles and
the particle-hole and particle-particle ladders.  The third and fourth
lines include all of the remaining fourth-order diagrams (the inclusion of the
second-order self energy into the dressed Green's function of the second-order
diagram produces another fourth-order contribution to the self energy, but
this is automatically included in the self-consistency step of the
conserving approximation).  The FLEX approximation consists of the summation
of all RPA bubbles, particle-hole ladders, and particle-particle ladders.
The self-energy has already been determined on the real axis by Menge
and M\"uller-Hartmann\cite{menge_muellerhartmann}.  The FLEX approximation
for the self energy includes all contributions thru third order in $U$
(the first line of Figure 8) but only a partial contribution of the
fourth-order and higher-order terms (the second line of Figure 8 plus the
higher-order terms).  An explicit formula for the electronic self energy
of the Hubbard model thru fourth order is given in the appendix.  The
corresponding formula for the FLEX approximation has been given
before\cite{baym,menge_muellerhartmann}.

The irreducible vertex functions are too cumbersome to represent
diagrammatically, but an explicit formula for the CDW vertex is given in the
appendix.  The corresponding formula for the FLEX approximation has already
been given\cite{baym}.  An additional simplification is usually made for the
FLEX approximation that neglects a large class of diagrams for the irreducible
vertices (the so-called Aslamazov-Larkin diagrams\cite{aslamazov_larkin}),
thereby including only those contributions to the irreducible vertex function
that can be represented by functions of the bare particle-hole or bare
particle-particle susceptibilities\cite{baym}.  This simplified FLEX
approximation will be denoted FLEX$^*$.

Since the Hubbard-model interaction is particle-hole symmetric, the half-filled
band corresponds to $\mu=0$, and the Green's functions are purely imaginary.
The odd-order contributions to the self energy all vanish
and each of the fourth-order contributions on a given line in Figure 8 are
identical\cite{yamada} (see the appendix).  Furthermore, it can be shown that
the most complicated contributions to the irreducible vertex function in
the CDW channel are odd under a change in the sign of the Matsubara
frequency, and can be neglected in calculating the maximum eigenvalue of
the scattering matrix, because only the even component of the irreducible
vertex function enters (see the appendix for details).

Note that since the self-energy is an odd function of $U$ at half-filling,
but the irreducible vertex function contains both even and odd powers of
$U$, the only difference between a truncated conserving approximation of
order $2n$ and of order $2n+1$ is that the irreducible vertex function
is larger for the odd-order approximation.  Therefore, we expect
that an even order approximation will underestimate the transition temperature
(in weak-coupling) and an odd-order approximation will overestimate $T_c$.

A comparison of the different approximation schemes is
given in Figures 9 and 10 for two
different values of $U$.  The second-order, third-order and FLEX approximations
all employ an energy cutoff of 256 positive Matsubara frequencies; the
fourth-order approximation uses 64 positive Matsubara frequencies.
In Figure 9~(a) the self-energy renormalization
function is plotted for the three different approximations at $U=-t^*$
and compared to the essentially exact IPT\cite{georges_kotliar}.
Note that the fourth-order approximation virtually reproduces the IPT
results, but that the FLEX approximation grossly overestimates the
self-energy even though the coupling strength is not too large.  In Figure
9~(b) the even component of one row of the irreducible vertex function
for the CDW channel is compared for $U=-t^*$.  All of the truncated
conserving approximations are in reasonable agreement with each other;
the FLEX approximation has the smallest magnitude at low Matsubara frequency.
The simplified FLEX$^*$ grossly overestimates the magnitude of the vertex
(in fact the FLEX$^*$ approximation produces the wrong qualitative behavior
of the vertex).  In general, the transition temperature calculated with the
simplified FLEX$^*$ will be a more accurate approximation to the
exact $T_c$ than that calculated with
the full FLEX, because the overestimation of the
self-energy will be compensated for by the overestimation of the vertex in
FLEX$^*$.  These results are similar to what White\cite{white} found for the
repulsive Anderson impurity model.

As the coupling strength is increased to $U=-2t^*$, the FLEX approximation
becomes a more accurate approximation
for the self energy than the truncated conserving
approximations [see Figure 10~(a)].  Clearly, the truncated conserving
approximation must be performed to a high order to accurately
reproduce the self energy in the strongly correlated regime.
The irreducible vertex function in the CDW channel is plotted in Figure
10~(b).  The different approximations no longer agree well with each other
indicating that the perturbation theory is breaking down.

The transition temperature for the CDW transition at half-filling in the
attractive Hubbard model is plotted in Figure 11 as a function of $U$.
The truncated conserving approximations are doing quite well in the
weak-coupling regime.  The odd-order approximation overestimates $T_c$,
while the even-order approximations underestimate $T_c$.  The
even-order approximations tend to be more accurate over a wider range of
$U$ than the odd-order approximation, but neither approximation properly
reproduces the turnover in $T_c$ as a function of $U$ as seen in the QMC
simulations\cite{jarrell}.  Note also that all truncated approximations
agree in the limit $U\rightarrow 0$, but that a first-order (RPA)
calculation will be off by a factor of three in the weak-coupling
limit\cite{vandongen,martin_flores,freericks_scalapino}.  The
FLEX approximation
does have the correct qualitative behavior of developing a peak in $T_c$
as a function of $U$ but the peak position and peak height are off by
about an order of magnitude.  The simplified FLEX$^*$ is, in general,
a more accurate approximation than the full FLEX, but becomes unstable
if $U$ is increased too far.  The FLEX$^*$ does not agree as well with
the QMC calculations (or with the other approximations) in the weak-coupling
limit because the irreducible vertex function does not include all of the
third-order contributions.

The reason why the truncated (even-order) conserving approximations do not
approximate the self energy (or the vertex) too accurately at moderate
coupling, but are good approximations for the transition temperature
at moderate coupling is most likely due to a cancellation of the effect
of an underestimation of the self energy by an underestimation of the
vertex in the calculation of $T_c$.  This probably explains why the $T_c$
curves do not turnover for the truncated approximations as well: the
turnover must be arising from self-energy effects that are being underestimated
here.

In summary, the truncated approximations tend to give better numerical
agreement than an approximation that tries to sum an infinite series of
diagrams (such as the FLEX).  One must go to very high order to see a peak
develop in $T_c$ as a function of $U$ and to have good quantitative
agreement with the QMC results.  It will be interesting to see if the
removal of the infinite summation of diagrams in the conserving approximation
produces an even better agreement with the QMC results (as was found for
the Anderson impurity model\cite{yamada}).  The approximation will no longer
be a conserving one, and will need to be generalized to move off of
half-filling, but should be even more accurate.  The fluctuation-exchange
approximation seems to be a poor approximation, and should not be tried
for the Holstein model, rather one should concentrate on generalizing
Yamada's analysis for instantaneous interactions to one for
retarded interactions to see whether or not one can improve upon the
accuracy in that case too.

\section{Conclusions}

Vertex corrections can be systematically incorporated into a weak-coupling
theory of electron-phonon interactions.  Expansions must be performed
to second order in the effective electron-electron interaction in order
to produce the correct behavior in the weak-coupling
limit\cite{vandongen,martin_flores}.
The miracle of
Migdal-Eliashberg theory\cite{migdal,eliashberg}
is that a first-order calculation suffices (with
dressed phonon propagators) in the small-phonon-frequency limit.
Vertex corrections enter to lowest order in the CDW channel, modifying
the interaction strength.  They enter to higher order in the SC channel,
and merely modify the prefactor of the weak-coupling $T_c$ equation.  This
robustness of the SC channel to the effects of vertex corrections
explains its remarkable
success for low temperature superconductors.  Nevertheless,
the effect of vertex corrections should be strong enough to be observable
in materials such as Ba$_{1-x}$K$_x$BiO$_3$ and the doped fullerenes.  It is
possible that the effects of vertex corrections can even be detected in
certain low-temperature superconductors such as Pb.

In general, vertex corrections will reduce the transition temperature,
however there is a small parameter regime at low electron density
where the vertex corrections actually cause an enhancement to the
superconducting transition temperature.  This occurs because the real
parts of the Green's functions are larger than the imaginary parts for
small imaginary frequency and low electron concentration.  At high
enough phonon frequency, or large enough electron density, the vertex
corrections will lower $T_c$.

Truncated conserving approximations appear to be better approximations than
infinite summation schemes such as the fluctuation-exchange
approximation\cite{baym}.
The electronic self energy, the irreducible vertex functions, and the
transition temperatures all appear to be better approximated by a truncated
conserving approximation.  The qualitative feature of the development of
a peak in the transition temperature as a function of the interaction strength
is, however, not reproduced by a truncated conserving approximation.  Perhaps a
completely truncated approximation (that is no longer conserving) will do even
better at approximating properties of interacting electronic systems in
infinite dimensions.  Yamada\cite{yamada,white} found this to be so for the
Anderson impurity model, and his techniques have been applied in infinite
dimensions\cite{georges_kotliar,georges_hub} to second order in $U$.  What is
needed is a way to generalize Yamada's work off of half filling for both the
self energy and the irreducible vertex functions.  Work in this direction is
currently in progress.

In conclusion, a weak-coupling conserving approximation has been carried out
for the attractive Holstein and Hubbard models that includes all effects
of vertex corrections and nonconstant density of states.  Agreement
with the exact solutions is found to be excellent at weak-coupling, but the
qualitative feature of developing a peak in $T_c$ as a function of the
interaction strength is not reproduced.  From this standpoint,
a weak-coupling theory is much more difficult to control than a strong-coupling
theory (perturbation theory in the kinetic energy).  Analytic expressions
for $T_c$ in the SC channel have been explicitly derived, and they indicate
that vertex corrections may be observable for some classes of low temperature
superconductors.

Future work will include an examination of the ordered phase, a study of the
effects of Coulomb repulsion, and a real materials calculation to look for the
effects of vertex corrections in low-temperature superconductors.

\acknowledgments

I would like to thank N. E. Bickers, T. Devereaux, M. Jarrell, E. Nicol,
and R. Scalettar
for many useful discussions.  I would especially like to thank D. Scalapino
for his continued interest in this problem and for numerous discussions.
This research is supported by the Office of Naval Research under Grant No.
N00014-93-1-0495.
Initial stages of this research were performed at the Institute for Theoretical
Physics in Santa Barbara and were supported in part by the NSF under Grants No.
PHY89-04035 and DMR92-25027.

\appendix
\section{Appendix}

The explicit formulas for the electronic self energy thru fourth order and
for the irreducible vertex function in the CDW channel are given here.
The diagrammatic expansion for the self energy is shown in Figure 8.
The first two lines include the contributions from the FLEX approximation
(truncated at fourth order) while the last two lines include the extra
fourth-order contributions.
The self energy is then expanded as
\begin{equation}
\Sigma (i\omega_n)\equiv\Sigma_n=\Sigma_n^{\rm FLEX}(4)+\Sigma_n^{\prime}(4)
\quad ,
\label{eq: hubbsigdef}
\end{equation}
with $\Sigma_n^{\rm FLEX}(4)$ the contributions to the self energy included in
the FLEX approximation\cite{baym} (but truncated to fourth order)
and $\Sigma_n^{\prime}(4)$ the additional
contributions to fourth order.
The truncated FLEX contributions are\cite{baym}
\begin{equation}
\Sigma_n^{\rm FLEX}(4)=TU\sum_lG_{n-l}[U\chi_l^{ph}+U^2\chi_l^{ph2}+
2U^3\chi_l^{ph3}]+TU\sum_lG_{-n-1+l}[U^2\chi_l^{pp2}-U^3\chi_l^{pp3}]\quad ,
\label{eq: sigmaflex}
\end{equation}
thru fourth order.  Here the bare particle-hole and particle-particle
susceptibilities are
\begin{equation}
\chi^{ph}(i\omega_l)=\chi_l^{ph}=-T\sum_rG_rG_{r+l}\quad ,\quad
\chi^{pp}(i\omega_l)=\chi_l^{pp}=T\sum_rG_rG_{-r-1+l}\quad .
\label{eq: chiph_chipp}
\end{equation}
The additional fourth-order contributions to the self energy are
\begin{eqnarray}
\Sigma_n^{\prime}(&4&)=T^2U^4\sum_{ll^{\prime}}G_{n-l}G_{n-l+l^{\prime}}
G_{n+l^{\prime}}\chi_l^{ph}\chi_{l^{\prime}}^{ph}
+2T^2U^4\sum_{ll^{\prime}}G_{-n-1+l}G_{-n-1+l-l^{\prime}}G_{n+l^{\prime}}
\chi_l^{pp}\chi_{2n+1-l+l^{\prime}}^{ph}\cr
&-&T^3U^4\sum_{rll^{\prime}}G_{n-l}G_rG_{r+l}G_{r+l+l^{\prime}}G_{r+l^{\prime}}
\chi_{l^{\prime}}^{ph}
-T^3U^4\sum_{rll^{\prime}}G_{n-l}G_{n-l^{\prime}}G_rG_{r-l+l^{\prime}}
G_{r+l^{\prime}}\chi_{n+r+1-l}^{pp}\cr
&+&T^3U^4\sum_{rll^{\prime}}G_{n-l}G_{n-l^{\prime}}G_rG_{r+l}G_{r+l^{\prime}}
\chi_{n-m-l-l^{\prime}}^{ph}\quad .
\label{eq: sigma4th}
\end{eqnarray}

At half filling the Green's functions are purely imaginary and satisfy
$G_{-n-1}=-G_n$.  Therefore, the particle-hole and particle-particle
susceptibilities in Eq.~(\ref{eq: chiph_chipp}) are equal.  It is easy to
show that the two third-order contributions to $\Sigma^{\rm FLEX}(4)$ vanish
in this case, and that the three fourth-order contributions are equal.
Similarly, the three terms that involve a double summation in
Eq.~(\ref{eq: sigma4th}) are equal, and so are the three triple summation terms
as shown by Yamada\cite{yamada}.

The irreducible vertex function in the CDW channel can also be determined.
The vertex function is broken up into its FLEX contributions and its
additional fourth order contributions
\begin{equation}
\Gamma_{mn}^{\rm CDW}=\Gamma_{mn}^{\rm FLEX}(4)+\Gamma_{mn}^{\prime}(4)\quad .
\label{eq: hubbgamma}
\end{equation}
The FLEX contributions thru fourth order are\cite{baym}
\begin{eqnarray}
\Gamma_{mn}^{\rm FLEX}(&4&)=U+U^2\chi_{m-n}^{ph}[2+U\chi_{m-n}^{ph}+
2U^2\chi_{m-n}^{ph2}]-U^2\chi_{m+n+1}^{pp}[1-U\chi_{m+n+1}^{pp}+
U^2\chi_{m+n+1}^{pp2}]\cr
&-&2TU^2\sum_rG_rG_{n-m+r}[U\chi_{m-r}^{ph}+3U^2\chi_{m-r}^{ph2}]
-2TU^2\sum_rG_rG_{m+n-r}[U\chi_{m-r}^{ph}+3U^2\chi_{m-r}^{ph2}]\cr
&-&2TU^2\sum_rG_rG_{m-n+r}[-2U\chi_{m+r+1}^{pp}+3U^2\chi_{m+r+1}^{pp2}]\quad ,
\label{eq: gammaflex}
\end{eqnarray}
and the additional fourth-order contributions are
\begin{eqnarray}
\Gamma_{mn}^{\prime}(&4&)=2TU^4\sum_rG_rG_{n-m+r}[\chi_{m-r}^{ph}
\chi_{n+r+1}^{pp}
+\chi_{m-n}^{ph}\chi_{m-r}^{ph}+\chi_{m-n}^{ph}\chi_{n+r+1}^{pp}]\cr
&+&TU^4\sum_rG_rG_{m+n-r}[2\chi_{m-r}^{ph}\chi_{m+n+1}^{pp}+\chi_{m-r}^{ph}
\chi_{n-r}^{ph}]\cr
&+&4T^2U^4\sum_{rs}G_rG_sG_{n-r+s}G_{n-m+s}[\chi_{m-r}^{ph}+\chi_{n-r}^{ph}]\cr
&-&2T^2U^4\sum_{rs}G_rG_sG_{m+n-s}[G_{n+r-s}\chi_{m-r}^{ph}+
G_{m+r-s}\chi_{n-r}^{ph}]\cr
&-&2T^2U^4\sum_{rs}G_rG_sG_{-m+r+s}G_{n-m+s}[\chi_{m-r}^{ph}+\chi_{n-r}^{ph}]\cr
&-&T^2U^4\sum_{rs}G_rG_s[G_{n-m+r}G_{n-m+s}-G_{m+n-r}G_{m+n-s}]\chi_{r-s}^{ph}
\cr
&-&2T^2U^4\sum_{rs}G_rG_sG_{m+r-s}G_{n+r-s}[\chi_{m+r+1}^{pp}+\chi_{n+r+1}^{pp}]
-T^2U^4\sum_{rs}G_rG_sG_{n-m+r}G_{m-n+s}\chi_{r+s+1}^{pp}\cr
&-&2T^2U^4\sum_{rs}G_rG_s[G_{m+r-s}G_{n-r+s}\chi_{m+r+1}^{pp}+G_{n+r-s}
G_{m-r+s}\chi_{n+r+1}^{pp}]\cr
&-&4T^3U^4\sum_{rst}G_sG_t{\rm Re}[G_rG_{r+s-t}G_{-m+r+s}G_{n+r-t}]\cr
&+&T^3U^4\sum_{rst}G_rG_sG_tG_{r+s-t}G_{n+s-t}[G_{m-r+t}+G_{m+r-t}]\quad .
\label{eq: gamma4th}
\end{eqnarray}
Note that the simplified FLEX$^*$ approximation does not include any of the
terms that involve explicit summations over Matsubara frequency [the second and
third lines in Eq.~(\ref{eq: gammaflex})].

By making the transformations $t\rightarrow r+s-t$, $r\rightarrow s$, and
$s\rightarrow r$ in the triple summation terms in Eq.~(\ref{eq: gamma4th})
and using the symmetry at half filling $G_{-n-1}=-G_n$, one can demonstrate
that
the triple summation terms are {\it odd} under $n\rightarrow -n-1$, and
do not contribute to the eigenvalue of the scattering matrix if the
eigenvector is even under $n\rightarrow -n-1$.  Therefore, the triple-summation
terms in Eq.~(\ref{eq: gamma4th}) may be neglected (this result has been
explicitly tested by calculating the eigenvalue of the scattering matrix
with and without the triple-summation terms and there was no effect on
the eigenvalue at half filling).

%\end{document}

\begin{figure}
  \caption{Dyson equations for the self energy of the Holstein model.
The thick solid lines denote {\it dressed} (electronic)
Green's function and the thin wavy lines denote the phonon propagator.
The self energy (expanded out to second order in a
conserving approximation) is depicted in (a)
and includes the Hartree and Fock contributions,
the second-order dressing of the phonon line and the lowest-order vertex
correction.  The self-consistent equation for the
self energy in Migdal-Eliashberg theory is shown in (b).  The thick
wavy line is the dressed
phonon propagator which satisfies the Dyson equation in (c).
 }
  \label{fig:1}
\end{figure}

\begin{figure}
\caption{The irreducible vertex functions in the CDW and SC channels.
The CDW irreducible vertex function for ME theory is shown in (a).  Note
that the phonon propagator is {\it bare} in (a).  The SC irreducible
vertex function for ME theory appears in (b).  Note that the phonon propagator
is dressed here.  The CDW irreducible vertex function for the second-order
conserving approximation is shown in (c).  Note that the vertex corrections
(exchange diagrams) modify the interaction {\it to lowest order} in the
CDW channel.  The SC irreducible vertex function for the second-order
conserving approximation is shown in (d).  The vertex corrections first
enter at second order in the SC channel.}
\label{fig:2}
\end{figure}

\begin{figure}
\caption{ Comparison of the ME theory (solid line) to
the second-order conserving approximation (dashed line)
for the Holstein model at half filling with phonon
frequency $\Omega=0.5t^*$, interaction strength $g=0.4t^*$, and temperature
$T=t^*/16$.  This example is generic for the weak-coupling limit.  In (a)
the self energy renormalization function $Z(i\omega_n)-1$ is plotted
against the Matsubara frequency and compared to the IPT (solid dots).  In (b)
the symmetric combination of the first
column of the irreducible vertex function in the CDW channel
is shown.  Note that the
second-order conserving approximation is clearly superior to ME theory in the
limit of weak coupling.}
\label{fig:3}
\end{figure}

\begin{figure}
\caption{Comparison of the ME theory (solid line) to
the second-order conserving approximation (dashed line)
for the Holstein model at half filling with phonon
frequency $\Omega=0.5t^*$, interaction strength $g=0.5t^*$, and temperature
$T=t^*/9$.  This example is generic for the transition region to the
strong-coupling limit.  The self-energy renormalization function (a) and
the irreducible vertex function in the CDW channel (b) are both
pictured.  In (a) the self energy is compared to the IPT (solid dots).
Note that in the limit
where the strong-coupling effects begin to manifest themselves, the ME theory
is
becoming a more accurate approximation, or, put in other words,
the total effect of vertex corrections is
{\it reduced} as the interaction strength increases.}
\label{fig:4}
\end{figure}

\begin{figure}
\caption{Transition temperature to the CDW-ordered state at half filling in
the Holstein model at an intermediate phonon frequency $(\Omega=0.5t^*)$.
The ME theory (solid line) is compared to the second-order conserving
approximation (dashed line) and the QMC results (solid dots).  Note that
the vertex corrections are very important in the CDW channel and that only
the second-order conserving approximation produces the correct result in
the weak-coupling limit.  ME theory does, however, display the correct
qualitative behavior of developing a peak in $T_c$ as a function of interaction
strength.}
\label{fig:5}
\end{figure}

\begin{figure}
\caption{Phase diagram of the Holstein model with $\Omega=0.5t^*$ at two
different coupling strengths $(g=0.4,0.5)$.  The solid dots are the QMC
solutions with CDW order, and the open triangles are the QMC results with
SC order.  The kinks in the solid (ME) and dashed (second-order conserving
approximation) lines occur at the CDW-SC phase boundaries.  In the
weak-coupling
limit $(g=0.4)$ the second-order conserving approximation is superior to the
ME theory and is quite accurate for the SC transition.  The difference
between the ME results and the second-order conserving approximation show
explicitly the lowest-order effects of vertex corrections upon the SC
transition
temperature.  The effect of vertex corrections is reduced as the filling is
reduced.  ME theory is quantitatively more accurate in determining $T_c$
for the stronger coupling strength $(g=0.5)$ but the second-order conserving
approximation is superior in determining the CDW-SC phase boundary.  Clearly
both approximations are failing at such a large value of the interaction
strength.}
\label{fig:6}
\end{figure}

\begin{figure}
\caption{Plot of the lowest-order effect of vertex corrections on the
SC transition temperature in the weak-coupling limit.  In (a) the
renormalization factor for the transition temperature calculated with vertex
corrections divided by the transition temperature calculated without vertex
corrections is plotted as a function of the phonon frequency.  Eight different
values of the electron density are plotted
($\rho_e=0.1,0.2,0.3,0.4,0.5,0.6,0.7,
1.0$).  Note that the vertex corrections always reduce $T_c$ in the high
phonon frequency limit, but that vertex corrections initially {\it enhance}
$T_c$ at low electron density $(\rho_e<0.2)$.  At half-filling (where the
Green's functions are purely imaginary) vertex corrections will reduce $T_c$
by a factor of two when $\Omega/W=0.03$.  In (b) the prefactor of the
weak-coupling $T_c$ formula is plotted against the phonon frequency for the
same eight values of the electron density.  The cutoff frequency has been
chosen to satisfy $\omega_c=0.6\Omega$.  Note that there is always an optimal
phonon frequency where the SC response will be the largest, and that this
optimal phonon frequency increases as the electron concentration decreases.}
\label{fig:7}
\end{figure}

\begin{figure}
\caption{Self energy diagrams for the Hubbard model thru fourth order.  The
thick solid lines denote the {\it dressed} electronic Green's functions,
and the thin dotted lines are the Coulomb interaction.  The first two lines
contain all of the FLEX contributions truncated to fourth order.  The last
two lines are the remaining fourth-order diagrams.  At half filling the
odd-order contributions to $\Sigma$ vanish, and each of the three
fourth-order contributions on the same line yield the same contribution
to $\Sigma$.}
\label{fig:8}
\end{figure}

\begin{figure}
\caption{Comparison of the different conserving approximations
for the Hubbard model at half filling in the limit of weak-coupling
($U=-t^*,T=t^*/20)$.  The second-order (dashed line), third-order (dotted
line),
and fourth-order (solid line) conserving approximations
are compared to the full FLEX (chain dotted line) and
the simplified FLEX$^*$ (chain triple dotted line).
In (a) the self-energy renormalization function is plotted
against Matsubara frequency and compared to the IPT (solid dots).
In (b) the even component of the first column
of the irreducible vertex function in the CDW channel is plotted.  Clearly
the fourth-order approximation is the best approximation in this limit.
The FLEX approximation grossly overestimates the self energy.  The
simplfied FLEX$^*$ compensates for this by overestimating
the vertex to produce a more accurate value for $T_c$.}
\label{fig:9}
\end{figure}

\begin{figure}
\caption{Same as in Figure 9, but with a stronger value of the coupling
$(U=-2t^*,T=t^*/8)$.  In this limit the FLEX approximation is superior for
the self energy, but the vertex does not appear to be
reproduced accurately by any approximation.}
\label{fig:10}
\end{figure}

\begin{figure}
\caption{Transition temperature to the CDW-ordered state in the Hubbard
model at half filling.  The second-order (solid line), third-order (dotted
line), and fourth-order (solid line) conserving approximations are
compared to the full FLEX (chain dotted line), the simplified FLEX$^*$
(chain triple dotted line), and the QMC results (solid dots).  Note that the
odd order approximations overestimate $T_c$, the even-order approximations
underestimate $T_c$, and that one has to go to very high order to reproduce
the peak in the transition temperature as a function of the interaction
strength.  The FLEX approximation displays the correct qualitative behavior of
developing a peak, but is off by an order of magnitude in the peak position
and height.  The simplified FLEX$^*$ yields a quantitatively more accurate
approximation, but is poorer in the limit of weak coupling because it
does not include all of the third-order contributions properly.}
\label{fig:11}
\end{figure}

\end{document}